\documentclass[aps,pre,amsmath,amsart,amsfonts,floatfix,superscriptaddress]{revtex4}
\usepackage{graphicx,graphics}
\usepackage[]{epsfig}
\usepackage[T1]{fontenc}

\usepackage[english, american]{babel}
\usepackage{amsmath}
\usepackage[utf8]{inputenc}
\usepackage{pstricks}
\usepackage{indentfirst}
\usepackage{hyperref}

\newcommand{\beq}{\begin{equation}}
\newcommand{\eeq}{\end{equation}}
\newcommand{\beqd}{\begin{displaymath}}
\newcommand{\eeqd}{\end{displaymath}}
\newcommand{\beqa}{\begin{eqnarray}}
\newcommand{\eeqa}{\end{eqnarray}}
\renewcommand\appendix{\par
  \setcounter{section}{0}
  \setcounter{subsection}{0}
  \setcounter{figure}{0}
  \setcounter{table}{0}
  \renewcommand\thesection{Appendix \Alph{section}}
  \renewcommand\thefigure{\Alph{section}\arabic{figure}}
  \renewcommand\thetable{\Alph{section}\arabic{table}}
}

\usepackage[a4paper,top=2cm,bottom=2.cm,left=1.7cm,right=1.7cm]{geometry}

\begin{document}

\title{The jamming transition in high dimension:\\
an analytical study of the TAP equations and the effective thermodynamic potential}

\author{Ada Altieri} 
\email{ada.altieri@uniroma1.it} 
\affiliation{LPTMS, CNRS, Univ. Paris-Sud, Universit\'e Paris-Saclay, 91405 Orsay, France}
\affiliation{Dipartimento di Fisica, Sapienza Universit\`a di Roma, Piazzale A. Moro 2, I-00185, Rome, Italy}
\author{Silvio Franz}
\affiliation{LPTMS, CNRS, Univ. Paris-Sud, Universit\'e Paris-Saclay, 91405 Orsay, France}
 \author{Giorgio Parisi}
\affiliation{Dipartimento di Fisica, Sapienza Universit\`a di Roma, Piazzale A. Moro 2, I-00185, Rome, Italy}
\affiliation{Nanotec-CNR, UOS Rome, Sapienza Universit\`a di Roma, Piazzale A. Moro 2, I-00185, Rome, Italy}
\affiliation{INFN-Sezione di Roma 1, Piazzale A. Moro 2, 00185, Rome}

\begin{abstract}
We present a parallel derivation of the Thouless-Anderson-Palmer (TAP) equations and of an effective potential for the \emph{negative} perceptron and soft sphere models in high dimension. Both models are continuous constrained satisfaction problems with a critical jamming transition characterized by the same exponents.
Our analysis reveals that a power expansion of the potential up to the second order constitutes a successful framework to approach the jamming points from the SAT phase (the region of the phase diagram where at least one configuration verifies all the constraints), where the ground-state energy is zero. 
An interesting outcome is that approaching the jamming line the effective thermodynamic potential has a logarithmic contribution, which turns out to be dominant in a proper scaling regime. Our approach is quite general and can be directly applied to other interesting models.
Finally we study the spectrum of small harmonic fluctuations in the SAT phase recovering the typical scaling $D(\omega) \sim \omega^2$ below the cutoff frequency but a different behavior characterized by a non-trivial exponent above it.
 \end{abstract}

\maketitle

\section{Introduction}

In the last years the study of glasses at low temperature has attracted significant interest, both from a theoretical and an experimental point of view \citep{Wolynes, Hansen-McDonalds, Torquato-Stillinger, Sperl-Majmudar,Liu-Nagel}. Much attention has concentrated on systems formed by an athermal assembly \footnote{Working at zero temperature is appropriate for granular systems and foams, where the energy of even small rearrangements of configurations is orders of magnitude greater than the thermal energy at room temperature. This picture, however, fails for molecular glasses, where the temperature is a control parameter.} of repulsive particles with finite-range interactions, where one can observe - upon increasing density - a \emph{jamming transition}, corresponding to a rigid arrangement of particles. This phenomenon displays extremely interesting features with both first and second order transition behaviors: it is characterized by a discontinuity (e.g. in the particle coordination number) and at the same time by power-law scalings with highly universal critical exponents.

In this context, jamming emerges as a fundamental theoretical paradigm for constructing a low-energy theory of glasses. Studying a glass former far below the glass transition is a difficult task, because of the emergence of a new critical transition associated with a fractal landscape (the \emph{Gardner transition}) \citep{{Charbonneau-Parisi-Urbani},{KPUZ_II},{Urbani-Biroli}}, an activated dynamics and strong heterogeneities.
A full theory further exploring and explaining these low energy excitations properties is still \emph{in fieri}.
Compared to ordinary solids with a long-range crystalline order, the spectrum of low-energy excitations in jammed materials exhibits several anomalies \citep{Wyart-Nagel-Witten, Xu-Liu}. A striking feature characterizing amorphous solids is the violation of the expected Debye law, showing a plateau above a cut-off frequency $\omega^{*}$\footnote{This cut-off frequency corresponds to the emergence of the so-called Boson peak.}. A related question concerns the properties of these normal modes, which are highly heterogeneous and resonant near $\omega^{*}$ and become quasi-localized upon decreasing the frequency \cite{Xu-Vitelli-Liu-Nagel}. This aspect has non-trivial implications even in the thermal conductivity and the specific heat.

Given these premises, it is crucial to achieve a better theoretical understanding at all levels, in order to efficiently analyze the jamming transition and to bridge the gap between different scenarios. Important progress has been achieved considering systems of soft spheres in the limit of infinite space dimensions. In this case glassy phases and jamming can be exactly analyzed. 
The replica method has been used to study the static properties of the aforementioned phases, allowing an analytical derivation of critical exponents that give account for the values observed in numerical simulations, independent of space dimensions \citep{Charbonneau-Parisi-Urbani}. The analysis has been extended to dynamics \citep{KPZ, KPUZ_II, CKPUZ} allowing to derive Mode-Coupling-like equations for the correlation functions \citep{KMZ}.
In this work we complete the picture through the computation of the TAP \cite{TAP} free energy, the effective potential of the system.
We present an analytical derivation based on the Plefka expansion \citep{Plefka} or Georges-Yedidia expansion \citep{Georges-Yedidia, Zamponi_notes} in two examples of disordered systems, the negative perceptron \citep{SimplestJamming} and the soft spheres. Note that our approach is equivalent to consider a $1/d$ expansion around the mean-field solution, where $d$ represents the number of spatial dimensions. In infinite dimension only a finite number of diagrams is needed to have the exact expression for thermodynamic quantities, such as the magnetization or the free energy. 
The procedure to derive the following mean-field formalism is formally similar to that explained in \citep{Zdeborova-Krzakala}, applied there to optimization and inference problems. 
One of the first derivations of the TAP equations for the perceptron goes back to \citep{Mezard} where the author, using the cavity method on a binary model with $ \xi_{i}^{\mu}=\pm 1 $, provides a computation of the number of patterns which can be stored in an optimal neural network intended as an associative Hopfield-like memory. Our main purpose is instead to deal with critical properties of amorphous systems at zero temperature, close to the jamming threshold. 
Most of our analysis will concentrate on the negative perceptron model, which has been shown to be in the same universality class of high-dimensional spheres.
The perceptron was introduced long ago in neural networks and machine learning, exploited for years as a linear signal classifier \citep{Gardner, Gardner-Derrida}. However, an alternative interpretation has been proposed in \citep{SimplestJamming}.

Both models, the perceptron and the soft spheres, belong to a class of systems where one defines $M$ {\it gaps} $h_\mu(x)$ $(\mu=1,...,M)$ as functions of the system configurations $x=\lbrace x_1,...,x_N \rbrace$,  and a soft-constraint Hamiltonian:
\begin{equation}
\mathcal{H}[x]=\frac{1}{2} \sum_{\mu=1}^{M} h_{\mu}^2(x) \theta(-h_\mu(x)) \ .
\label{Hamiltonians}
\end{equation}

In the spheres $\mu \rightarrow (\alpha, \beta)$ codes for the pair of particles and $h_{\alpha \beta}(x)=\vert x_\alpha- x_\beta \vert -D$, where $x_\alpha$ and $x_\beta$ are the particle positions and $D$ is the diameter. In the following, we will look at arrangements of $N-1$ spheres on the surface of a $N$ dimensional sphere of radius $R$, this leads to a simpler analysis  than spheres in the  Euclidean space, but it is equivalent to it in the limit $D/R\to 0$. There are different possible high dimensional 
limits of the model, the natural one \citep{KPZ} consists in taking $R,M\to \infty$, $N\to \infty$ in this order, for fixed reduced packing fraction $\hat{\phi}=2^N\frac{M}{N}(D/2R)^{N-1}$, a second one consists in sending $N\to\infty$  $R=\sqrt{N}$ and $M=\alpha N$. In the regime where each particle interacts with $O(N)$ other particles, the limits commute and the two regimes can be smoothly connected. 

In the perceptron case one defines random gaps $h_{\mu}(\textbf{x})=\frac{\xi^{\mu} \cdot x}{\sqrt{N}} -\sigma$. The vectors $\xi^{\mu}$ are i.i.d quenched random variables with zero mean and unit variance. Also in this case we consider a spherical model, where $\sum_{i=1}^N x_i^2=N$.  For positive $\sigma$ the model is the usual perceptron classifier used in machine learning and defines a convex optimization problem, for negative $\sigma$ the model is non-convex and can be interpreted as the problem of a single $\emph{dynamical}$ sphere in a background of random obstacles $\xi^{\mu}$. In the following we will be interested in the latter case which is in the same universality class of spheres.  The interesting regime for the model is when $N$ and $M$ go to infinity for fixed ratio $\alpha=M/N$.

At zero temperature one seeks the minima of the Hamiltonian (\ref{Hamiltonians}). 
By varying the number $M$ of terms in the Hamiltonian one passes from a satisfiable region (SAT phase, with at least one configuration in agreement with all the requested constraints $h_\mu(x)>0$, called hard sphere side of the transition in jamming literature) to an unsatisfiable one (UNSAT phase, where the constraints cannot all be verified simultaneously, called soft sphere side of the transition). This SAT/UNSAT transition coincides with the equilibrium jamming transition. As a first example of application of the TAP formalism we choose to study the properties of the zero temperature SAT phase close to the transition. In fact in the UNSAT phase, the zero temperature free energy coincide with the energy. The UNSAT phase minima are isolated and their properties can be studied directly. The spectrum of these minima has been studied in \citep{FPUZ} recovering the typical glassy features above and at jamming. In the SAT phase conversely one has a zero energy manifold and the free energy measures its entropy.  Generically the energy is flat around minima and the Hessian is trivially zero.  However, the effective potential, \emph{i.e.} the free energy as a function of average particle positions in these regions is not flat.  Our analysis allows us to derive in the framework of exactly solvable models important properties of the jamming transition first found by Wyart and collaborators in the framework of approximate theories in finite dimension: (1) the form of the effective potential close to jamming is logarithmic in the gaps \citep{Wyart, DeGiuli-Wyart14, DeGiuli-Wyart}; (2) the spectrum of free energy fluctuations close to jamming displays non-trivial singularities \citep{DeGiuli-Wyart14, DeGiuli-Wyart}.

The organization of the paper is the following: in Section II we briefly give the main definitions to introduce an effective potential that will be treated in detail in Section III in the perceptron model and also in sphere models, where we discuss two different dimensional limits.
In Section IV we derive the leading behavior of the effective potential near jamming which exhibits a logarithmic contribution in the SAT phase and, using these results, we finally study in Section V the spectrum of small excitations.

\section{Definitions and main results}
We aim to compute the effective potential as a function of the mean particle position and to do this we start from the definition:
\begin{equation}
\label{uno}
e^{G({\bf m})}=e^{\sum\limits_{i=1}^{N} m_i u_i} \int d\textbf{x} \; e^{-\beta H[\textbf{x}]-\sum\limits_{i=1}^{N} x_i u_i} = e^{\bf{m\cdot u} + K[u]}
\end{equation}
evaluated at the point $u$ such that $\textbf{m} + \nabla_{\textbf{u}} K(\textbf{u})=0$.  As well known, $G[m]$ represents a coarse-grained free energy  where we integrate fast degrees of freedom. 
In order to compute $G(m)$ we found convenient to write a more general form of the potential to include generalized forces. Defining gap variables $h_\mu$ and enforcing 
$h_\mu=h_\mu(x)$ in the partition function through conjugate variables $i\hat{h}_\mu$ we rewrite (\ref{uno}) 
\begin{eqnarray}
  \label{eq:enforce}
e^{G({\bf m})}=e^{{\bf m\cdot u}} \int d\textbf{x} d{\bf h} d{\bf \hat{h}} \; e^{-\beta H[\textbf{h}]-{\bf x\cdot u} -\sum_{\mu} i \hat{h}_\mu( h_\mu(x)-h_\mu)} .
\end{eqnarray}
The variables $i\hat{h}_\mu$ are conjugated to the gaps and can be interpreted as generalized forces. We therefore introduce a more general effective potential, function also of generalized forces,  defined from
\begin{eqnarray}
  \label{eq:due}
  e^{\Gamma({\bf m},\bf {f})}=e^{{\bf m\cdot u}+\sum_\mu f_\mu v_\mu} \int d\textbf{x} d{\bf h} d{\bf \hat{h}} \; e^{-\beta H[\textbf{h}]-{\bf x\cdot u}-\sum_{\mu} i \hat{h}_\mu( h_\mu(x)-h_\mu+v_\mu)}
=e^{J({\bf u},{\bf v}) +{\bf m \cdot u} +{\bf f \cdot v}} 
\end{eqnarray}
with 
\begin{eqnarray}
  \label{eq:tre}
  \frac{\partial J}{\partial u_i}= \frac{\partial J}{\partial v_\mu}=0\;\;\;\; \forall i,\mu.
\end{eqnarray}
We write explicitly 
\begin{eqnarray}
&&\Gamma(m,f)=\sum_{i} m_i u_i +\sum_{\mu}f_\mu v_{\mu}-\log \int_{x,h_\mu,\hat{h}_\mu}  e^{S_\eta(x_i,h_\mu,\hat{h}_\mu)} \ , \\
&&S_\eta(x,h_\mu,\hat{h}_\mu)=   u_i \cdot x_i +  i v_\mu \cdot \hat{h}_\mu -\lambda(x_i^2-1) +\frac{\beta}{2} h_\mu^2 \theta(-h_\mu)-i \hat{h}_\mu (h_\mu-\eta h_\mu(x))-\frac{b}{2} (\hat{h}_\mu^2-\tilde{r}) \ .
\label{pot}
\end{eqnarray}
Here $\eta$ is a formal parameter that we introduce for later convenience and should be set to one at the end of the computation, $\tilde{r}$ is defined as $\tilde{r}=-\frac 1 M \sum_\mu \hat{h}_\mu^2$ and has to be fixed at the end of the computation by an extremum condition, $b$ is a Lagrange multiplier that enforces that condition. 
From the definition of $\Gamma$ we clearly have:
\begin{equation}
G(m)= \Gamma(m,f) \hspace{0.7cm} \text{evaluated in} \hspace{0.7cm} \frac{\partial \Gamma(m,f)} {\partial f}=0 \ .
\end{equation}
To derive the TAP free energy we perform a Plefka expansion of the term $H_{eff}(\textbf {x})=i\hat{h}_\mu h_\mu(x)$ in the action. This amounts formally to performing a Taylor expansion in $\eta$, which is equivalent to an expansion in $1/N$. The leading terms are obtained truncating the expansion to the terms of order $\eta^2$. 

Useful compact notations are reported in the following, identifying respectively the Edwards-Anderson parameter (a.k.a. \emph{self-overlap}) and the first two moments of the average variable $i \hat{h}_\mu $:
\begin{equation}
q=\frac{1}{N}\sum_{i} m_i^2 \ , \hspace{0.6cm} 
r=-\frac{1}{\alpha N} \sum_{\mu=1}^{M} f_{\mu}^2  \ ,\hspace{0.6cm}
\tilde{r}=-\frac{1}{\alpha N} \sum_{\mu=1}^{M} \langle\hat{h}^2_{\mu} \rangle \ .
\end{equation}
From the stationary condition $\frac{\partial G}{\partial \tilde{r}}=0$, we get the value of the Lagrange multiplier $b$, namely $b=1-q$. 

The Plefka expansion requires the computation of the following terms:
\begin{equation}
\frac{\partial \Gamma}{\partial \eta}= - \langle H_{eff} \rangle 
\end{equation}
\begin{equation}
\frac{\partial^2 \Gamma}{\partial \eta^2}=- \left \{ \langle H_{eff}^2 \rangle -\langle H_{eff} \rangle^2-\langle H_{eff} \left[ \sum_i \frac{\partial u_i}{\partial \eta} (x_i -m_i)+\sum_\mu \frac{\partial v_\mu}{\partial \eta}(i\hat{h}_\mu-f_\mu) \right] \rangle \right \} \ .
\label{second_deriv}
\end{equation}
As we said before, we can neglect terms of order $\eta^3$ and higher that correspond to 
vanishing contributions in the limit $N\to\infty$ \citep{Georges-Yedidia, Crisanti-Sommers}.

\section{Effective potential for the negative perceptron}

Let us focus on the perceptron model.
The first term appearing in the power expansion is expressed as the average effective Hamiltonian depending on the conjugated variables $h_\mu$, $i\hat{h}_\mu$:
\begin{equation}
\langle H_{eff}\rangle=  \sum_{i,\mu} \frac{ \xi_{i}^{\mu}  m_i f_{\mu}}{\sqrt{N}}
\end{equation}
As far as the second order term is concerned, in principle one should consider in Eq. (\ref{second_deriv}) several mixing terms. We have checked that only those with equal indexes ($\mu=\nu$, $i=j$) provide a non vanishing contribution (see \ref{App:AppendixA}). 
The general TAP expansion of the potential is:
\begin{equation}
\begin{split}
\Gamma(m,f)=&\sum_{i} \phi(m_i)+\sum_{\mu} \Phi(f_{\mu})+\left.\frac{\partial \Gamma}{\partial \eta} \right|_{\eta=1}\eta+
\left.\frac{1}{2}\frac{\partial^2 \Gamma}{\partial \eta^2} \right|_{\eta=1} \eta^2+\mathcal{O}(\eta^3)=\\
=&\sum_{i} \phi(m_i)+\sum_{\mu} \Phi(f_{\mu})-\sum_{i,\mu} \frac{\xi_{i}^{\mu}m_i f_{\mu}}{\sqrt{N}}
+\frac{\alpha N}{2}  (\tilde{r}-r)(1-q) 
\end{split}
\label{effective_pot}
\end{equation}
where 
\begin{equation}
\phi(m)= \min_{u} \left[ m u -\log \int dx e^{-\lambda (x^2-1)+ u x} \right] \ .
\end{equation}
Using the integral representation of the delta function enforcing the spherical constraint, the integral can be evaluated for large $N$ via a saddle point, implying:
\begin{equation}
\sum_i \phi(m_i) \approx -\frac{N}{2} \log(1-q) \ .
\end{equation}
The latter term in Eq. (\ref{effective_pot}) plays the role of an \emph{Onsager reaction term}, describing the fluctuations between $\langle \hat{h}_\mu^2\rangle$ and $\langle \hat{h}_\mu \rangle^2$ \footnote{In the Sherrigton-Kirkpatrick model, a pivotal example of disordered fully connected system, the analogous expression for the free energy is: $-\beta F[m]=\sum_{i} s(m_i) +\frac{\beta}{2} \sum\limits_{i \neq j} J_{ij} m_i m_j +\frac{N \beta^2}{4}(1-q)^2$, where the Onsager term is proportional only to $(1-q)^2$. Here we find a more complicate situation with a dependence even on the first two moments of the force.}.
Hence the resulting expression of the effective potential reads:
\begin{equation}
\begin{boxed}{
\Gamma(m,f)=-\frac{N}{2}\log (1-q)+\sum_{\mu} \Phi(f_{\mu})-\sum_{i,\mu} \frac{ \xi_{i}^{\mu} f_{\mu}m_i}{\sqrt{N}} 
+\frac{\alpha N}{2} \left[ (\tilde{r}-r)(1-q) \right] 
\label{free_energy}
}
\end{boxed}
\end{equation}
where the second term has the following form:
\begin{equation} 
\begin{boxed}{
\Phi(f)={{\min_{v}}} \left[f  v-\log \int \frac{dh d\hat{h}}{2\pi} e^{\frac{\beta}{2} h^2 \theta(-h)-i \hat{h}(h+\sigma)+ i v \hat{h} -\frac{b}{2}(\hat{h}^2-\tilde{r}) } \right] 
}
\end{boxed}
\end{equation}
We remind that at the saddle-point $b=1-q$.
Note that by integrating over $\hat{h}$, the expression above leads to a simple Gaussian integral in the gaps.

The main saddle-point equations which enable to characterize our model are:
\begin{equation}
\frac{\partial \Gamma}{\partial m_i}=0  \hspace{0.5cm} \Rightarrow \hspace{0.5cm}
m_i \left(\frac{1}{1-q}-\alpha(\tilde{r}-r)\right)= \sum_{\mu} \frac{\xi_{i}^{\mu} f_{\mu}}{\sqrt{N}}
\label{saddle-p_m}
\end{equation}
\begin{equation}
\frac{\partial \Gamma}{\partial f_{\mu}}=\Phi^{'}(f_\mu)-\sum_{i}\frac{ \xi_{i}^{\mu}mi}{\sqrt{N}}+(1-q)f_{\mu}=0 
\label{stat_g}
\end{equation}

\subsection{Generalization to soft sphere models}

The analysis of the perceptron suggests an immediate generalization to the sphere problem  if we consider $M=\alpha N$ particles with positions $x_\alpha$, ($\alpha=1,...,M$)  on the $N$-dimensional sphere with radius $R=\sqrt{N}$, $x_\alpha^2=N$, where the gap variables are written as: 
\begin{equation}
h_{\alpha \beta} = \frac{x_{\alpha} \cdot x_{\beta}}{\sqrt{N}} -\sigma \ .
\label{gap_spheres}
\end{equation}
Notice that this is not the natural scaling that would lead to the infinite-dimensional limit of Euclidean space, which consists in taking $R\to\infty$ before sending the dimension of the space to infinity. The crucial point is to work in a regime where each particle effectively interact with $O(N)$ other particles. Both of the mentioned regimes have this property. Despite we derive the TAP free energy with the former scaling it will be valid for the latter as well in a suitable limit $\alpha, \sigma\to\infty$. 

The effective Hamiltonian we use for the Plefka expansion now reads:
\begin{equation}
H_{eff}= i \sum_{\langle \alpha, \beta \rangle} \frac{\hat{h}_{\alpha \beta} x_\alpha \cdot x_\beta}{\sqrt{N}} \ .
\end{equation}
Using the same treatment applied for the perceptron, we obtain:
\begin{equation}
\frac{\partial \Gamma}{\partial \eta}= \langle H_{eff} \rangle =-\sum_{\langle \alpha \beta \rangle} \frac{f_{\alpha \beta} m_\alpha \cdot m_{\beta}}{\sqrt{N}} \ ,
\end{equation}
\begin{equation}
\frac{\partial^2 \Gamma}{\partial \eta^2}=- \left [ \langle H_{eff}^2 \rangle -\langle H_{eff} \rangle^2-
\langle H_{eff} \sum_{i} (s_i-m_i)  \frac{\partial \langle H_{eff} \rangle }{\partial m_i} \rangle \right]
\label{second_derivative1}
\end{equation}
where we have generically indicated with $s_i$ both types of variables, positions and contact forces. 
We normalize the parameters of the model in this way:
\begin{equation}
q=\frac{1}{NM} \sum_{i,\alpha} (m_i^{\alpha})^2 =\frac{1}{M}\sum\limits_{\alpha=1}^{M} \frac{m^{\alpha} \cdot m^{\alpha}}{N} \ , \hspace{0.8cm} 
r=-\frac{1}{MN} \sum_{\alpha \beta} f_{\alpha \beta}^2  \ ,\hspace{0.8cm}
\tilde{r}=-\frac{1}{MN} \sum_{\alpha \beta} \langle\hat{h}^2_{\alpha \beta} \rangle \ .
\end{equation}
anticipating that we shall consider the regime where each particle interacts with $O (N)$ particles and therefore there are $O(N)$ forces $f_{\alpha \beta}$ for each sphere.

The first contribution in Eq. (\ref{second_derivative1}) reads:\begin{equation}
\langle H^2 \rangle -\langle H \rangle^2= -\frac{1}{N} \sum \left( \langle \hat{h}_{\alpha \beta} \hat{h}_{\gamma \delta} \rangle \langle x_i^{\alpha} x_i^{\beta} x_j^{\gamma} x_j^{\delta} \rangle \right)_{c}= -MN (\tilde{r}-r q^2 )
\end{equation}
where the only surviving terms are those with $\alpha \beta = \gamma \delta$.
In the same way we can write the second term and the resulting expression for the free energy is:
\begin{equation}
\Gamma(m,f)=-\frac{MN}{2} \log (1-q)+\sum_{\alpha \beta} \Phi(f_{\alpha \beta})-\sum_{\alpha \beta} \frac{f_{\alpha \beta} m_\alpha \cdot m_\beta}{\sqrt{N}} + \frac{ MN}{2}\left[(\tilde{r}-rq^2)+ r q(1-q) + q^2 (\tilde{r}-r) \right]\ ,
\label{potential_spheres}
\end{equation}
where $\Phi(f)$ is the same function as for the perceptron case, \emph{i.e.}:
\begin{equation} 
\Phi(f)={{\min_{\tilde{v}}}} \left[f  \tilde{v}-\log \int \frac{dh d\hat{h}}{2\pi} e^{\frac{\beta}{2} h^2 \theta(-h)-i \hat{h}(h+\sigma)+ i v \hat{h} -\frac{\tilde{b}}{2}(\hat{h}^2-\tilde{r}) } \right] 
\end{equation}
As before, $\tilde{v}$ is the Lagrange multiplier associated with the force and $\tilde{b}$ is another multiplier enforcing the average value of $\hat{h}_{\alpha \beta}^2$.
We have checked that third-order $O(\eta^3)$ terms in Eq. (\ref{potential_spheres}) provide a subleading contribution with respect to the first two. Since only the first two moments are dominant in the expansion (that implies a Gaussian distribution of the random component $\xi^{\alpha \beta}$), one can see that the same expansion would work in an equivalent disordered model where the scalar products in the definition of 
$h_{\alpha,\beta}$ are substituted by random couplings:
\begin{eqnarray}
  \label{eq:couplings}
  h_{\alpha,\beta}(x)=\frac{1}{N}\sum_{ij} \xi^{\alpha \beta}_{ij} x_i^\alpha x_j^\beta-\sigma
\end{eqnarray}
where the $\xi_{ij}^{\alpha \beta}$ are choosen as independent, variance one Gaussian variables. This is a manifestation of the equivalence of self-generated disorder models and models with quenched disorder in high dimension \citep{MPR, Bouchaud-Mezard, Cugliandolo-Kurchan,Mari-Kurchan}.

\section{Logarithmic interaction near random close packing density}

In this Section we analize the effective potential in the SAT phase close to jamming. For simplicity we use the percepton free energy, but the same analysis and results could be obtained for the spheres. We get then an exact derivation within our models of the logarithmic interaction derived by Brito and Wyart \citep{Wyart}, studying the microscopic cause of rigidity of three dimensional hard-sphere glasses. 
In the SAT phase the function $\Phi$ appearing in TAP free energy reduces to:
\begin{equation}
\Phi(f)=f v -\log H{ \left( \frac{ \sigma -v}{\sqrt{1-q}} \right)}
\end{equation}
where the function $H(x) =\int_{x}^{\infty} \frac{1}{\sqrt{2\pi}}e^{-\frac{t^2}{2}} dt $ is the complementary error function. 
The forces $f_\mu$ are the partial derivative of $\Phi$ with respect to the Lagrange multiplier $v_\mu$:
\begin{equation}
f_\mu= - \frac{H'\left(\frac{\sigma-v_\mu}{\sqrt{1-q}}\right)}{\sqrt{1-q} H\left( \frac{\sigma-v_\mu}{\sqrt{1-q}} \right) }   \ .
\end{equation}
Close to jamming $q \rightarrow 1$ and the argument $\frac{\sigma -v_\mu}{\sqrt{1-q}} \gg 1 $. Considering the asymptotic expansion of the function $
 H(x) \approx  \frac{e^{-\frac{x^2}{2}}}{x \sqrt{2\pi}} $, we get the following expression for the potential:
\begin{equation}
\Phi(f)= f  v -\log H{ \left( \frac{\sigma- v}{\sqrt{1-q}} \right)} \approx  f \cdot v +\Theta(\sigma-v) \left[ \frac{(\sigma-v)^2}{2(1-q)} +\log \left( \frac{\sigma -v}{\sqrt{1-q}} \right) \right] \ .
\end{equation}
At this point we can simplify the expression of the generalized forces $f_\mu$, constrained over the value $\sigma-v_\mu$:
\begin{equation}
f_\mu= \Theta(\sigma-v_\mu) \left( \frac{\sigma-v_\mu}{1-q} +\frac{1}{\sigma-v_\mu} \right) \ .
\end{equation}
Then we use Eq. (\ref{stat_g}), which we rewrite as:
\begin{equation}
\sigma-v_\mu= -h_\mu({\bf m}) +(1-q)f_\mu \ ,
\end{equation}
where we posed $h_\mu({\bf m})=\frac{\xi^{\mu} \cdot m}{\sqrt{N}} -\sigma$.
Putting all these results together, we get that to the leading order the average gap is inversely proportional to the generalized contact force:
\begin{equation}
h_\mu({\bf m})=\frac{1-q}{\sigma -v_\mu} \approx \frac{1}{f_\mu} \ .
\end{equation}
Taking into account that the expansion of the complementary error function $H(x)$ only holds for large values $\frac{\sigma-v_\mu}{\sqrt{1-q}} \gg 1$, the previous one can be rewritten in the following form: $\frac{ h_\mu(\bf{m}) }{\sqrt{1-q}} \ll 1$.
Clearly we have also: $\frac{h_\mu(\bf{m})}{1-q} \ll \frac{1}{h_\mu(\bf{m})}$.

By eliminating $f_\mu$ from the previous equations and expressing all the quantities in terms of $h_\mu(\textbf{m})$, we see that to the leading order the effective potential reads:
\begin{equation}
\begin{split}
G(m)\simeq  & -\frac{N}{2} \log(1-q)+\sum_{\mu} \Theta(\sigma -v_\mu)\left[ \frac{ h_\mu(\textbf{m})  ^2}{2(1-q)} +\log \left(  -h_\mu(\textbf{m}) +\frac{1-q}{ h_\mu(\textbf{m}) } \right) -\frac{1}{2} \log(1-q)+... \right] \\
\simeq & -\sum_{\mu} \theta (h_\mu(\textbf{m}))  \log \left( \frac{h_\mu(\textbf{m})} {1-q} \right) 
\end{split}
\end{equation}
We have to take care of the first two terms in parentheses: depending on the appropriate scaling regime, one can observe either a quadratic or a logarithmic behavior. 
Neglecting irrelevant numerical prefactors and the first entropic term, the leading contribution turns out to be logarithmic in the gap.
Therefore, instead of describing the model through a hard-core potential which involves several difficulties, we can replace it with a smooth logarithmic interaction dominating when approaching the jamming line from the SAT phase. 
We have verified that the same behavior also holds for hard spheres in agreement with the argument explained in \citep{Wyart, DeGiuli-Wyart14, DeGiuli-Wyart}.

\section{Spectrum of small fluctuations}

Starting from Eq. (\ref{free_energy}) and differentiating with respect to $m_i$, we can directly compute the Hessian of the potential, where we assume to neglect the projector term $\frac{2m_im_j}{N(1-q)^2}$ and all the other $1/N$ order terms. If included in the computation, the projector splits off a single isolated eigenvalue from the continuous band \citep{aspelmeier-bray-moore}.
At the end we get (more details in \ref{App:spectrum}):
\begin{equation}
\mathcal{M}_{ij} \equiv \frac{d^2 G}{d m_i d m_j}=\delta_{ij} \left[\frac{1}{1-q}-\alpha (\tilde{r}-r) \right]-\sum_{\mu}\frac{\xi_{i}^{\mu}\xi_{j}^{\mu}}{N}
\frac{1}{\Phi^{''}(f_\mu)+(1-q)} \ .
\label{hessian}
\end{equation}
The diagonal term $\zeta= \frac{1}{1-q}-\alpha (\tilde{r}-r) $ gives only a shift in the spectrum, while the $\mu$-dependent terms multiplying $\xi_{i}^{\mu}\xi_{j}^{\mu}$ are the stiffness parameters defined as 
\begin{eqnarray}
  \label{eq:quattro}
k_\mu=-\frac{\partial f_\mu}{\partial h_\mu}=-\frac{1}{\Phi^{''}(f_\mu) +(1-q)}.  
\end{eqnarray}
Note that close to jamming $q\to 1$ and the Hessian must be rescaled by  $(1-q)$ in order to have finite eigenvalues. This rescaling comes from the definition of the potential which corresponds indeed to a sort of $\emph{vibrational entropy}$, describing the volume of space around a given metastable state. 

The average resolvent $R(z)$ associated to the Hessian is:
\begin{equation}
R^{(N)}(z)=\frac{1}{N} \langle Tr \left( z \textbf{I} -\textbf{M} \right)^{-1} \rangle 
\end{equation}
with $z=\lambda-i\epsilon$, in terms of a regularized parameter $\epsilon>0$. 
As usual the spectrum $\rho(\lambda)$ can be computed from a limiting procedure on the resolvent:\begin{equation}
\rho^{(N)}(\lambda)=\frac{1}{\pi} \lim_{\epsilon \rightarrow 0^{+}} \Im (R^{(N)}(z)) \ .
\end{equation}

The key point is that the stiffness terms $k_\mu$ and the random part $\xi_{i}^{\mu} \xi_{j}^{\mu}$ can be considered uncorrelated for large $N$, because each $k_\mu$ is a sum of a huge number of patterns and depends only marginally on each of them. We get therefore for the Hessian the usual form of a modified Wishart matrix and as a consequence the resolvent satisfies the equation:
\begin{equation}
\lambda+\zeta=\frac{1}{R}+\frac{1}{N}\sum_{\mu=1}^{M} \frac{k_\mu}{1-k_\mu R } =\frac{1}{R}+F^{'}(R)
\label{resolvent}
\end{equation}
where $F(R)=-\frac{1}{N} \sum_{\mu}^{M} \log(1-k_\mu R)$.
A way to derive the equation above might consist in writing the expectation value of the resolvent in field theory where its cumulants give the correlators. These correlators, once introduced a loop insertion operator, satisfy loop equations, namely the \emph{Schwinger-Dyson equations} or \emph{Pastur equations} in mathematical jargon \citep{Borot-Nadal}.
\subsection{Spectrum in the SAT phase close to Jamming}
Equation (\ref{resolvent}) cannot be solved exactly. However, we can identify the leading behaviors of the distribution of eigenvalues at low frequency close to jamming. 
The lower edge of the spectrum $\lambda_0$, which happens for $R=R_0$, is identified by the equations:
\begin{eqnarray}
&&\lambda_0+\zeta=\frac{1}{R_0}+F^{'}(R_0)
\\
&& -\frac{1}{R_0^2}+F''(R_0) =0
\label{edge}
\end{eqnarray}
which express the fact that close to $R_0$ there is no real linear solution $\lambda-\lambda_0\propto R-R_0$ and 
\begin{equation}
\lambda-\lambda_0 = \left( \frac{1}{R_0^3}+F'''(R_0)\right) \left(R-R_0 \right)^2.
\end{equation}
This, for non-zero $\zeta$, leads immediately to a square-root behavior of the spectrum close to the edge $\rho(\lambda) \sim \sqrt{\lambda-\lambda_0}$. From (\ref{edge}) one can observe that the condition that the spectral gap $\lambda_0$ is positive coincides with the condition that the so-called replicon eigenvalue in the replica solution of the model is positive. This is known to be positive in liquid (Replica Symmetric) phase and in stable (1RSB) phases, while it vanishes in (full-RSB) spin-glasses or marginal Gardner glass phases. It is known \citep{SimplestJamming, FPUZ, FPSUZ}  that close to the jamming line the perceptron is in a Gardner phase, the replicon is zero and therefore the spectral gap vanishes. Moreover, when jamming is approached $\zeta \rightarrow 0$ and $R \rightarrow \infty$. Clearly the square-root behavior of the spectrum is modified for $\lambda \sim \zeta$. In order to better understand the typical behavior we need to study the following integral over the stiffness probability distribution $p(k)=(1-A)\delta(k)+AP(k)$ where $A$ is the fraction of non zero forces, which tends to $1$ at jamming. We need then:
\begin{equation}
F'(R)= \frac{1}{R} A\int_{0}^{\infty} \text{d}k \hspace{0.05cm} \text{P}(k) \frac{k R}{1+kR} 
\label{Fprime}
\end{equation}
evaluated for large $R$. Clearly, the first term in the expansion is $F'[R]\approx A/R$, for $A\to 1$.  This term cancels with the first in (\ref{resolvent}) and we need to evaluate the next term in the expansion
\begin{equation}
F'(R)= \frac{1}{R} A\left[ 1- \int_{0}^{ \infty} \text{d}k \hspace{0.05cm} \text{P}(k) \frac{1}{1+kR}\right] \ .
\label{Fprimedue}
\end{equation}
The last term depends critically on the form $P(k)$ close to the origin, which we are going to investigate. We recall that $k_\mu=-[\Phi''(f_\mu)+(1-q)]^{-1}.$ For $q\to 1$, to the leading order, $\Phi''(f_\mu)^{-1}=-\frac{1}{1-q}+\frac{1}{(\sigma-v)^2}$, so that using the results of the previous section $k_\mu \approx f_\mu^2$. One needs then the distribution of forces, which, quoting from the replica solution \citep{SimplestJamming, FPUZ} behaves as a power $P(f)\sim f^\theta$ with a non-trivial exponent $\theta  \sim 0.42311$. 
We find that $ \int_{0}^{\infty} \text{d}k \hspace{0.05cm} \text{P}(k) \frac  1 k \approx \int_{0}^{\infty} \text{d}f \hspace{0.05cm} \text{P}(f)\frac{1}{f^2}$ is divergent, and therefore to the leading order
\begin{eqnarray}
  \label{eq:leading}
\int df\;  \text{P}(f) \frac{1}{1+f^2R}\approx \frac{C}{R^{\frac{1+\theta}{2}}} 
\end{eqnarray}
where $C$ is a constant. 
The equation for the resolvent can be written to the leading order as:
\begin{equation}
\lambda+ \zeta = \frac{1-A}{R}  +\frac{C}{R^{\frac{3+\theta}{2}}} \ ,
\end{equation}
which leads to
\begin{eqnarray}
  \label{eq:spettro}
  \rho(\lambda)\sim \left\{ 
\begin{array}{ll}
\vspace{0.5cm}
Const. \sqrt{\lambda} & \hspace{0.3cm} \lambda\ll \zeta \\
 \frac{Const. }{(\zeta+\lambda)^{\frac{2}{3+\theta}}} & \hspace{0.3cm} \lambda \ge \zeta
\end{array}
\right.
\end{eqnarray}
with the two regimes being well interpolated by a form $\rho(\lambda)\approx Const. \frac{\sqrt{\lambda }}{(\zeta+\lambda)^{\frac{7+\theta}{2(3+\theta)}}}.$
These results coincide with  the one found by \citep{DeGiuli-Wyart14} once expressed in terms of the  density of states
 $D(\omega)=\rho(\lambda) \frac{d \lambda}{d \omega}$, with $\lambda= \omega^2$.
Notice that the behavior of the spectral density differs from the one of the Hessian of the Hamiltonian on the UNSAT 
side of the transition, which has the form $\rho_{UNSAT}(\lambda)\propto \frac{\sqrt{\lambda}}{\lambda+\zeta}$ with $\zeta\to 0$ at the transition \citep{FPUZ, CCPPZ}. This discontinuity should not surprise too much, given the way we have taken the zero temperature limit.  In the SAT phase, any increase of the energy away from zero is forbidden,  vibrations are purely entropic and their amplitude  is proportional to the cage size $\sqrt{1-q}$.

\section{Conclusions}

The perceptron is a useful model for glassy systems close to jamming. It allows simplified derivations of general properties of more high-dimensional sphere models. 
In this paper we have derived the effective potential in the perceptron and, exploiting the gained insight, generalized it to the sphere model. We have studied the effective potential in the SAT phase close to jamming, exhibiting the logarithmic form found 
in \citep{Wyart} for finite dimensional spheres. In the same regime we also studied the spectrum of entropic fluctuations finding that the spectrum depends on the exponent of the force distribution at jamming. In perspective we will study the finite properties of the model, that allows interpolating between the SAT and the UNSAT regimes.

\subsection*{Acknowledgements} 
We would like to thank Francesco Zamponi and Pierfrancesco Urbani for very useful discussions. We also thank Patrick Charbonneau for his suggestion for future developments. 
This work was supported by a grant from the Simons Foundation ($\#454949$  to Giorgio Parisi and $\#454941$ to Silvio Franz).

\appendix
\section{Leading contribution in the second order term of the free energy}
\label{App:AppendixA}

We have shown above that one of the relevant term to compute in the Plefka-like expansion is the second derivative of the potential:
\begin{equation}
\frac{\partial^2 \Gamma}{\partial \eta^2}=- \left \{ \langle H^2 \rangle -\langle H \rangle^2-\langle H \left[ \sum_i \frac{\partial u_i}{\partial \eta} (x_i -m_i)+\sum_\mu \frac{\partial v_\mu}{\partial \eta}(i\hat{h}_\mu-f_\mu) \right] \rangle \right \} =\alpha N(\tilde{r}-r)(1-q) \ .
\label{second_der}
\end{equation}
We want to give a brief sketch of the reason why off-diagonal terms - if $(ij,\mu \nu)$ are all different or there are only two equal indexes - do not contribute to the general form of the potential.
Let us focus on the first term in Eq. (\ref{second_der}):
\begin{equation}
\langle H^2 \rangle -\langle H \rangle^2=\sum_{i j,\mu \nu}\frac{\xi_{i}^{\mu}\xi_{j}^{\nu}}{N} \langle x_i x_j  i \hat{h}_{\mu} i \hat{h}_{\nu} \rangle_{c} \ .
\label{connected_energy}
\end{equation}
The sum in Eq. (\ref{connected_energy}) receives contributions from the following possible combinations of indexes:
\begin{itemize}
\item $i \neq j, \mu \neq \nu$: the connected function is trivially zero \\
\item $i = j, \mu \neq \nu$: $\left( \langle x_i^2 \rangle -m_i^2 \right) f_\mu f_\nu$ \\ 
\item $i \neq j, \mu = \nu$: $\left(\langle (i\hat{h}_{\mu})^2 \rangle - \langle i\hat{h}_{\mu} \rangle^2 \right) \langle x_i \rangle \langle x_j \rangle=\left(\langle (i\hat{h}_{\mu})^2 \rangle - \langle i\hat{h}_{\mu} \rangle^2 \right)m_i m_j$  \\
\item $i = j, \mu = \nu$: $\left( \langle x_i^2 \rangle \langle (i\hat{h}_\mu)^2 \rangle - \langle x_i \rangle^2 \langle i\hat{h}_\mu \rangle ^2 \right)$
\end{itemize}
Concerning the second part:
\begin{equation}
\begin{split}
\langle H \left[ \sum_{j} \frac{\partial u_j}{\partial \eta} (x_j -m_j)+ \sum_{\nu} \frac{\partial v_\nu}{\partial \eta}(i\hat{h}_\nu-f_\nu) \right] \rangle &=
\langle \sum_{i,\mu}\frac{ \xi_{i}^{\mu}x_i i\hat{h}_\mu}{\sqrt{N}} \sum_{j, \nu} \frac{\xi_{j}^{\nu} f_{\nu}}{\sqrt{N}} (x_j-m_j)\rangle+ \langle \sum_{i,\mu}\frac{ \xi_{i}^{\mu}x_i i\hat{h}_\mu}{\sqrt{N}} \sum_{j, \nu} \frac{\xi_{j}^{\nu} m_j}{\sqrt{N}} (i\hat{h}_\nu-f_\nu)\rangle\\
&=\frac{1}{N} \sum_{ij,\mu \nu} \left[ \xi_{i}^{\mu} \xi_{j}^{\nu} f_\mu f_\nu (\langle x_i x_j \rangle) -m_i m_j )+  \xi_{i}^{\mu} \xi_{j}^{\nu} m_i m_j (\langle i\hat{h}_\mu i\hat{h}_\nu \rangle  -f_\mu f_\nu) \right]
\end{split}
\end{equation}
we give some examples, neglecting for the moment the contribution of the patterns:
\begin{itemize}
\item $i = j, \mu \neq \nu$: $\left( \langle x_i^2 \rangle -m_i^2 \right) f_\mu f_\nu$ it cancels with the second term in the list above \\ 
\item $i \neq j, \mu = \nu$: $\left(\langle (i\hat{h}_{\mu})^2 \rangle - \langle i\hat{h}_{\mu} \rangle^2 \right)m_i m_j$ it cancels with the third term \\
\item $i = j, \mu = \nu$: $ \alpha r(1-q)+\alpha q (\tilde{r}-r) $ 
\end{itemize}
The only relevant case occurs when $i=j$ \& $\mu=\nu$, otherwise these correlation functions are zero either because they are totally disconnected (if $i \neq j$,$\mu \neq \nu$) or because they sum up to zero combined with another term of opposite sign. 

\hspace{3cm}

\section{Detailed computation of the Hessian matrix}
\label{App:spectrum}
To derive the Hessian matrix of the potential $G(m)$
we need to express the function $f_\mu$ in terms of the average variable $m_i$.

\begin{equation}
\frac{d^2 \Gamma}{d m_i dm_j}=\frac{\partial^2 \Gamma}{ \partial m_i \partial m_j} + \sum_{\mu=1}^{M} \frac{\partial ^2 \Gamma} {\partial f_\mu m_j} \frac{\partial f_\mu}{\partial m_i} +\sum_{\mu=1}^{M} \frac{\partial ^2 \Gamma} {\partial f_\mu m_i} \frac{\partial f_\mu}{\partial m_j} +\sum_{\mu, \nu}^{M} \frac{\partial ^2 \Gamma} {\partial f_\mu \partial f_\nu} \frac{\partial f_\mu}{\partial m_i}  \frac{\partial f_\nu}{\partial m_j} 
\end{equation}
An important remark is the stationary condition of the potential:
\begin{equation}
\frac{\partial \Gamma}{\partial f_{\mu}}=\Phi^{'}(f_\mu)-\sum_{i}\frac{ \xi_{i}^{\mu}mi}{\sqrt{N}}+(1-q)f_{\mu}=0 
\end{equation}
from which we have:
\begin{equation}
\left[\Phi^{''}(f_\mu)+(1-q) \right] \frac{\partial f_\mu}{\partial m_i} -\frac{\xi_{i}^{\mu}}{\sqrt{N}}=0 \ .
\end{equation}
This implies:
\begin{equation}
\frac{\partial ^2 \Gamma} {\partial f_\mu^2 } \frac{\partial f_\mu}{\partial m_i} + \frac{\partial^2 \Gamma}{\partial f_{\mu} \partial m_i}=0
\end{equation}
and as a consequence the last two terms in the Hessian cancel with each other.
Moreover, noticing that the partial derivative is simply:
\begin{equation}
\frac{\partial^2 \Gamma}{\partial f_\mu \partial m_j}=-\frac{\xi_{j}^{\mu}}{\sqrt{N}}
\end{equation}
and 
\begin{equation}
\frac{\partial ^2 \Gamma}{ \partial f_\mu f_\nu} =\delta_{\mu\nu} \left[\Phi''(f_\mu) +(1-q) \right] \ ,
\hspace{1cm} 
\frac{\partial ^2 \Gamma}{ \partial m_i \partial m_j} =\delta_{ij} \left[\frac{1}{1-q} -\alpha(\tilde{r}-r) \right ] +\frac{2m_im_j}{N(1-q)^2}
\end{equation}
(as explained before, the last term can be neglected giving a subleading contribution in $1/N$), the resulting expression for the Hessian reads:
\begin{equation}
\begin{split}
\mathcal{M}_{ij} \equiv \frac{d^2 G}{d m_i d m_j}=& \delta_{ij} \left[\frac{1}{1-q} -\alpha(\tilde{r}-r) \right] -\frac{2}{N} \sum_{\mu=1}^{M} \left[ \Phi''(f_\mu)+(1-q) \right] ^{-1} \xi_{i}^{\mu} \xi_{j}^{\mu}+\frac{1}{N} \sum_{\mu=1}^{M} \left[ \Phi''(f_\mu)+(1-q) \right] ^{-1} \xi_{i}^{\mu} \xi_{j}^{\mu}\\
=& \delta_{ij} \left[\frac{1}{1-q} -\alpha(\tilde{r}-r) \right] -\frac{1}{N} \sum_{\mu=1}^{M} \left[ \Phi''(f_\mu)+(1-q) \right] ^{-1} \xi_{i}^{\mu} \xi_{j}^{\mu}
\end{split}
\end{equation}


\end{document}